%Paper: hep-ph/9308211
%From: JPHALYO@WISWIC.WEIZMANN.AC.IL
%Date: Tue, 3 Aug 1993 13:32:45 GMT
%Date (revised): Sun, 12 Sep 1993 10:57:03 GMT
%Date (revised): Sun, 12 Sep 1993 14:29:13 GMT

%       This paper needs the macro packages phyzzx.tex and tables.tex
%
%
%
\input phys
\tolerance=1000
\sequentialequations
\def\rl{\rightline}

\def\t1{{\tilde 1}}

\def\AEF{A.E. Faraggi}

\def\NPB#1#2#3{Nucl. Phys. B {\bf#1} (19#2) #3}
\def\PLB#1#2#3{Phys. Lett. B {\bf#1} (19#2) #3}

\def\PRL#1#2#3{Phys. Rev. Lett. {\bf#1} (19#2) #3}

\REF\CH{H. Y. Cheng, Phys. Rep. {\bf 158} (1988) 1.}
\REF\PQ{R. Peccei and H. Quinn, Phys. Rev. Lett. {\bf 38} (1977) 1440; Phys.
Rev. D {\bf 16} (1977) 1791.}
\REF\WW{S. Weinberg, \PRL{40}{78}{223}; F. Wilcek, \PRL{40}{78}{279}.}
\REF\DFS{M. Dine, W. Fischler and M. Srednicki, \PLB{104}{81}{199}.}
\REF\GSW{M. Green, J. Schwarz and E. Witten, Superstring Theory, 2 vols.,
Cambridge University Press, 1987.}
\REF\W{E. Witten, \PLB{149}{84}{351}.}
\REF\KIM{K. Choi and J. Kim, \PLB{154}{84}{393}; \PLB{165}{85}{71}.}
\REF\ROSS{G. Lazarides, C. Panagiotakopoulos and Q. Shafi, \PRL{56}{86}{432};
J. A. Casas and G. G. Ross, \PLB{192}{87}{119}.}
\REF\IBA{L. E. Ibanez and D. Lust, \PLB{267}{91}{51}.}
\REF\NAN{J. L. Lopez and D. V. Nanopoulos, \PLB{245}{90}{111}.}
\REF\SLM{\AEF, \NPB{387}{92}{289}.}
\REF\EU{\AEF, \PLB{278}{92}{131}; \PLB{274}{92}{47}.}
\REF\FFF{I. Antoniadis, C. Bachas and C. Kounnas,
Nucl. Phys. B {\bf 289}
(1987) 87; I. Antoniadis and C. Bachas, Nucl. Phys. B {\bf 298} (1988)
586; H. Kawai, D.C. Lewellen and S.H.-H. Tye, Phys. Rev. Lett. {\bf57} (1986)
1832; Phys. Rev. D {\bf 34} (1986) 3794;
Nucl. Phys. B {\bf 288} (1987) 1.}
\REF\DSW{M. Dine, N. Seiberg and E. Witten, \NPB{289}{87}{585}.}
\REF\CKMM{A. E. Faraggi and E. Halyo, \PLB{307}{93}{305}; WIS-93/35/APR-PH.}

\singlespace
\rl{WIS--93/66/JUL--PH}
\rl{\today}
\rl{T}
\normalspace
\smallskip
\titlestyle{\bf{Can the Axions of Standard--like Superstring
 Models Solve the Strong CP Problem?}}
\author{Edi Halyo{\footnote*{e--mail address: jphalyo@weizmann.bitnet}}}
\smallskip
\centerline {Department of Physics, Weizmann Institute of Science}
\centerline {Rehovot 76100, Israel}
\vskip 6.0 cm

\titlestyle{\bf ABSTRACT}

We find that there are three axions in standard--like superstring
models in the four dimensional free fermionic formulation. These axions
are either harmful or very heavy. Therefore,
they cannot solve the strong CP problem. We show that this is a general result
in superstring models with chiral generations from the $Z_2$ twisted sectors
which use a $Z_4$ twist.

%\nopagenumbers
\pagenumber 0
\singlespace
\vskip 0.5cm
\endpage
\normalspace

\centerline{\bf 1. Introduction}

One of the unsolved problems in the Standard model is the strong CP problem
[\CH] i.e. why $\bar \theta=\theta_{QCD}+\theta_{quark}<10^{-9}$?
The most elegant solution to the strong CP problem is the
Peccei--Quinn (PQ) mechanism [\PQ]. In the PQ mechanism there is a color
anomalous
global $U(1)$ current which is a symmetry of the Lagrangian at the classical
level. During electro--weak symmetry breaking this $U(1)_{PQ}$ is also
spontaneously broken and, as a result, there is an axion [\WW]. In order to
make the
axion phenomenologically acceptable, one introduces a $SU(2)_L$ singlet scalar
which has a nonzero PQ charge and obtains a very large VEV. The axion, which
then becomes ``invisible",  gets a small mass only from instanton effects in
QCD [\DFS]. In this way, one turns $\bar \theta$ into a dynamical variable (the
axion) with a minimum at $\bar \theta=0$.

In general, the definition of the PQ symmetry and the $SU(2)_L$ singlet scalar
introduced are quite arbitrary. The only purpose for their introduction is to
solve the strong CP problem. One would hope that these essential features of
the PQ mechanism  will arise automatically from a more fundamental theory which
reduces to the Standard model at low energies.
To date, the most complete theory at very high energies (i.e. $M_P$) is
superstring theory [\GSW]. If superstring theory describes nature, it has to
give phenomenologically acceptable results. In particular, it has to
solve the strong CP problem. Therefore, it is important to check whether a
PQ--like symmetry and the
required $SU(2)_L$ singlet scalars arise naturally in superstring--induced
models.

It is well known that there is a model independent axion in superstring models
compactified on Calabi--Yau manifolds [\W]. This axion necessarily couples to
the hidden sector gauge groups as well as the observable gauge groups. As a
result, it gets large a mass when the hidden gauge groups become strongly
interacting [\KIM]. Therefore, it cannot solve the strong CP problem.
Another possibility is the existence of an approximate PQ--like global symmetry
as a result of discrete symmetries [\ROSS].
A PQ--like symmetry is also seen in models based on orbifold compactifications
 [\IBA].
Here, if one requires the $N=1$ supergravity model to be target space modular
invariant and also Kahler, the colored fermions must transform under an
associated chiral $U(1)$ symmetry. This chiral $U(1)$ is
identified as a PQ symmetry. In this case, the corresponding axion might
solve the strong CP problem under some assumptions.
For superstrings in the four dimensional free fermionic formulation, in flipped
$SU(5)$ models, it has been shown that there are additional axions [\NAN].
Unfortunately, these are either very massive or harmful axions which cannot
solve the strong CP problem.

In this paper, we investigate axions in standard--like superstring models
in the four dimensional free fermionic formulation [\SLM,\EU]. We find that
there are three axions in the model. The first is the model independent
axion which is present in all superstring models. The second is the Goldstone
boson of an anomalous $U(1)_A$ and the third corresponds to a non--anomalous,
approximate, global $U(1)$ which has color and hidden sector gauge group
anomalies. We show that none of these axions can solve the strong CP
problem. This is because they are either very massive or harmful. They are
massive if they couple to the non--Abelian hidden gauge groups and these
condense when they become strong.
They are harmful if the VEV of the corresponding scalar with the nonzero PQ
charge (which gives the axion constant) is too large. The only way out of
this situation is to find a color
anomalous global $U(1)$, which is broken at an acceptable scale and has no
anomalies with respect to the non--Abelian hidden gauge groups. We show
that, in general, this is not possible in models which have chiral generations
from the $Z_2$ twisted sectors and which use a $Z_4$ twist.

\bigskip
\centerline{\bf 2. The superstring standard--like models}

The superstring standard--like models are constructed in the four
dimensional free fermionic formulation [\FFF].
The models are generated by a basis of eight boundary condition vectors
for all world--sheet fermions. The first five vectors in the basis
consist of the NAHE set $\{{\bf 1},S,b_1,b_2,b_3\}$ [\SLM].
The standard--like models are constructed by adding three additional
vectors to the NAHE set [\SLM,\EU].
These vectors and the choice of generalized GSO projection coefficients for our
model are given in Table 1 [\EU].
The observable and hidden gauge groups after application
of the generalized GSO projections are
$SU(3)_C\times U(1)_C\times
 SU(2)_L\times U(1)_L\times U(1)^6${\footnote*{
$U(1)_C={3\over 2}U(1)_{B-L}$ and
$U(1)_L=2U(1)_{T_{3_R}}$.}}
and $SU(5)_H\times SU(3)_H\times U(1)^2$, respectively.
The weak hypercharge and the combination orthogonal to it are given by
$U(1)_Y={1\over 3}U(1)_C + {1\over 2}U(1)_L$ and $U(1)_{Z^\prime}=
U(1)_C - U(1)_L$ respectively.
The six gauge $U(1)_r$ correspond to the right--moving world--sheet
currents ${\bar\eta}^j_{1\over2}{{\bar\eta}^{j^*}}_{1\over2}$
($j=1,2,3$) and ${{\bar y}^3{\bar y}^6,{\bar y}^1{\bar\omega}^5,
{\bar\omega}^2{\bar\omega}^4}$.
For every right--moving $U(1)_r$ gauge symmetry there is
a left--moving global $U(1)_\ell$ symmetry. The six $U(1)_\ell$
correspond to the left--handed world--sheet currents
$\chi^{12}$, $\chi^{34}$, $\chi^{56}$, $y^3y^6$, $y^1\omega^5$ and
$\omega^2\omega^4$.

A general property of the free fermionic models, which are based on the
NAHE set and use a $Z_4$ twist to break the gauge symmetry from $SO(2n)$ to
$SU(n) \times U(1)$,
is the presence of the sectors $b_j$ and $b_j+2\gamma+(I)$ (where $I=1+b_1+b_2
+b_3$) in the massless spectrum. The sectors $b_j$ produce the chiral
generations and the sectors $b_j+2\gamma+(I)$ produce vector representations
of the hidden gauge groups that are $SO(10)$ singlets.
We will argue that the $Z_4$ twist and its consequences mentioned above are
the essential features behind the fact that all axions are either harmful or
massive in these models.

The full massless spectrum was presented in Ref. [\EU].
Here we list only the states that are relevant for our purposes.
The sectors $b_{1,2,3}$  produce three $SO(10)$ chiral generations,
$G_\alpha=e_{L_\alpha}^c+u_{L_\alpha}^c+N_{L_\alpha}^c+d_{L_\alpha}^c+
Q_\alpha+L_\alpha$ $(\alpha=1,\cdots,3)$ with the correct
$SU(3)_C\times U(1)_C\times SU(2)_L\times U(1)_L$ charges.
%(b) The ${S+b_1+b_2+\alpha+\beta}$ sector gives the $SU(2)_L$ doublet
%%$h_{45}$,
%the $SU(3)_C$ triplet $D_{45}$ and the singlets $\Phi_{45}, \Phi^{\pm}_1,
%\Phi^{\pm}_2, \Phi^{\pm}_3$ (and their conjugates ${\bar h}_{45}$, etc.).
%
%(c) The Neveu--Schwarz $O$ sector gives, in addition to  the graviton,
%dilaton, antisymmetric tensor and spin 1 gauge bosons,  scalar
%electroweak doublets, $h_1, h_2, h_3$, singlets, $\Phi_{12}, \Phi_{13},
%\Phi_{23}$ (and their conjugates ${\bar h}_1$, etc.)and three singlet states,
%$\xi_{1,2,3}$, that are neutral under all the U(1) symmetries.
The sectors $b_i+2\gamma+(I)$ give states that are
$SU(3)_C\times SU(2)_L\times {U(1)_L}\times {U(1)_C}$
singlets and transform as $5$, ${\bar 5}$ and $3$, ${\bar 3}$
under the hidden $SU(5)_H$ and $SU(3)_H$ gauge groups, respectively.
States from the sectors $b_j$ and $b_j+2\gamma$ have nonzero charges under
$U(1)_{\ell_j}$, $U(1)_{\ell_{j+3}}$, $U(1)_{r_j}$ and $U(1)_{r_{j+3}}$ only.
In addition, there are $SO(10)$ singlet states coming from the
Neveu--Schwarz and $S+b_1+b_2+\alpha+\beta$ sectors.

The model contains an anomalous $U(1)$ gauge symmetry (which is given by
$U_A=2U_1+2U_2+2U_3-U_4-U_5-U_6$) with $Tr(Q_A)=180$ [\SLM,\EU].
The anomalous $U(1)_A$ induces a D term so that the
corresponding D contraint for supersymmetry at $M_P$ becomes [\DSW]
$$D_A=\sum_k Q^A_k \vert \chi_k \vert^2={{-g^2 e^{\phi_D}}\over{192 \pi^2}}
Tr(Q_A) \eqno(1)$$
In order to satisfy the F and D contraints and preserve supersymmetry at
$M_P$, a number of $SO(10)$ singlet scalars will
obtain VEVs. In general all local and global $U(1)$s will be spontaneously
broken [\SLM,\EU].

\bigskip
\centerline{\bf 3. The axions of the model}

The first axion, $a_1$, is the model independent axion which is present in all
superstring models [\W]. In the Neveu--Schwarz sector, in four dimensions,
there is always
an antisymmetric tensor $B_{\mu \nu}$ which has one physical degree of freedom.
$B_{\mu \nu}$ is equivalent to an axion by duality, i.e.
$$H_{\mu \nu \lambda}=*da_1={1\over {6M}}\epsilon_{\mu \nu \lambda \sigma}
\partial^{\sigma}a_1 \eqno(2)$$
where $H$ is the $B$ strength tensor given by $H=dB+\omega_L-\omega_{YM}$ and
$\omega_L$ and $\omega_{YM}$ are the gravitational and Yang--Mills
Chern--Simons
forms. From the Bianchi identity
$dH=-(TrF^2-TrR^2)$ one gets the axion couplings
$$\partial^{\mu}\partial_{\mu} a_1=-{1\over M}(TrF \tilde F-Tr R \tilde R)
\eqno(3)$$
where the axion constant is given by $f_{a_1}\sim 10^{16}~GeV$ [\KIM]. Note
that $a_1$
couples to the hidden sector gauge groups in addition to the observable groups.

The second axion, $a_2$, arises as the Goldstone boson of an anomalous global
$U(1)_A$ in the theory [\NAN]. As mentioned before, in standard--like
superstring models,
there is an anomalous local $U(1)_A$. This apparent anomaly at the field theory
level is cancelled by a string induced Green-Shwarz counterterm, $BF$, in four
dimensions ($F$ is the $U(1)_A$ field strength) [\DSW]. This leads to a term of
the form
$\partial^{\mu}a_1 A_{\mu}$ which breaks $U(1)_A$ and gives mass ($\sim M$) to
the gauge boson. The model independent axion, $a_1$, becomes the longitudinal
component of the massive gauge boson.
Still, a (anomalous) global $U(1)_A$ remains as a symmetry of the model at the
classical level. In addition to the total anomaly, $Tr U(1)_A=180$, $U(1)_A$
is also anomalous in the QCD and hidden $SU(5)_H$ and $SU(3)_H$ sectors:
$$TrU(1)_AH_1H_1=15, \qquad TrU(1)_AH_2H_2=10, \qquad TrU(1)_AGG=15 \eqno(4)$$
where $H_1,H_2$ and $G$ are the hidden $SU(3)_H,SU(5)_H$ and $SU(3)_C$ color
gauge groups respectively.

{}From the D constraint for $U(1)_A$, we see that $U(1)_A$ is spontaneously
broken.
In fact, generically, it is broken by the VEVs of a large number of scalars.
$U(1)_A$ is a candidate for PQ symmetry since it is color anomalous and also
spontaneously broken by $SU(2)_L$ singlet scalars at a high scale.

Under $U(1)_A$, the scalars $\phi_i$ transform as
$\phi_i \to \phi_i^{\prime}=\phi_i e^{iq_i \alpha}$
where $\alpha$ is the parameter of the transformation and $q_i$ are the
$U(1)_A$ charges of the scalars. After spontaneous symmetry breakdown, one
can write $\phi_i=(\rho_i+v_i)e^{i\chi_i/ v_i}$ where $v_i$ are the scalar
VEVs. Then the axion $a_2$ is given by
$a_2=(1/ f_{a_2})\sum_i v_i q_i \chi_i$
with the axion constant $f_{a_2}^2=\sum_i v_i^2 q_i^2$. The axion, $a_2$ is a
linear combination of
the phases of the scalars which break $U(1)_A$. $a_2$ and $f_{a_2}$ are
dominated by the phase and VEV of the scalar with the largest VEV respectively.
Due to the $U(1)_A$ anomaly in the hidden sector (both $SU(5)_H$ and
$SU(3)_H$),
$a_2$ also couples to the hidden sector gauge bosons.

Due to their couplings to $A_{\mu}$, $a_1$ and $a_2$ mix when $v_i \not=0$. The
mixing is of the form $(f_{a_2}/M)\partial_\mu a_1 \partial^\mu a_2$. The
physical axions are given by $a={1\over \sqrt 2} (a_1+a_2)$ and $a^{\prime}=
{1\over \sqrt 2} (a_1-a_2)$. The physical axions' interactions with the gauge
bosons are given by
$${\cal L}_{int}=(H_1 \tilde H_1+G \tilde G)\left({a\over f_a} + {a^{\prime}
\over f_{a^{\prime}}}\right)+ H_2 \tilde H_2 \left({\alpha a \over f_a} +
{\beta a^{\prime} \over f_{a^{\prime}}}\right) \eqno(5)$$
where $H_1,H_2$ and $G$ are the $SU(3)_H,SU(5)_H$ and $SU(3)_C$ field strength
tensors and tilde symbolizes the dual. The physical axion constants are
$$f_a={{f_{a_2}+15f_{a_1}}\over {\sqrt 2 f_{a_1}f_{a_2}}} \qquad
  f_{a^{\prime}}={{f_{a_2}-15f_{a_1}}\over {\sqrt 2 f_{a_1}f_{a_2}}} \eqno(6)$$
and $\alpha=(f_{a_2}+10f_{a_1})/(f_{a_2}+15f_{a_1})$ and
$\beta=(f_{a_2}-10f_{a_1})/(f_{a_2}-15f_{a_1})$.

In standard--like superstring models, the scalar VEVs resulting from the F and
D
constraints are at the scale ${M/10} \sim 10^{17}~GeV$ ( where $M=M_P/{2 \sqrt
{8 \pi}}$) as fixed by the
anomaly term in Eq. (1). In general, all scalars obtain VEVs which break
$U(1)_A$ spontaneously. Therefore, generically, $a_2$ is a linear combination
of a large number of scalar phases with an axion constant $f_{a_2} \sim
{M/10} \sim 10^{17}~GeV$ fixed by the VEVs. From Eq. (6) one finds $f_a \sim
f_{a^{\prime}} \sim 10^{17}~GeV$.

There is an additional axion, in standard--like superstring models, which
arises
as follows. At the trilevel of the superpotential, there are six global
$U(1)$s, without an
overall anomaly, which correspond to left--handed world--sheet currents.
These global symmetries are spontaneously broken by the $SO(10)$ singlet VEVs.
Higher order non--renormalizable corrections to the superpotential induce
terms which break these symmetries explicitly. One has approximate global
symmetries if these terms are small because they appear at high orders.
Of the six $U(1)$s, only $U(1)_6$, which corresponds to the world--sheet
current $\omega_2 \omega_4$ has QCD and hidden $SU(5)_H,SU(3)_H$ anomalies. The
other five $U(1)$s do not have QCD anomalies and therefore are irrelevant for
the strong CP problem. The only fields with $U(1)_6$ charge are {$Q_3$,
$u^c_3$, $d^c_3$, $T_3$, $\bar T_3$, $V_3$, $\bar V_3$} ($L_3$, $e^c_3$,
$N_3$ and $SU(3)_C$, $SU(5)_H$, $SU(3)_H$ neutral hidden sector states which
are irrelevant for our purposes) with the charges [\SLM,\EU]
$Q_6(u^c_3,T_3,\bar T_3)=-Q_6(Q_3,d^c_3,V_3,\bar V_3)={1 \over 2}$.
$U(1)_6$ is explicitly broken only by the effective light quark and lepton
mass terms
$Q_3d_3$, $L_3e_3$ and by the mixing terms $Q_3u_2$, $Q_2u_3$, $Q_1u_3$,
$Q_3u_1$, $Q_3d_2$,
$Q_2d_3$, $Q_1d_3$, $Q_3d_1$ . Since these are small [\CKMM], $U(1)_6$ is an
approximate global symmetry. For the anomalies one has
$$TrU(1)_6 GG=-1,\qquad TrU(1)_6 H_1H_1=-1, \qquad TrU(1)_6 H_2H_2=1 \eqno(7)$$

So, $U(1)_6$ is a candidate for PQ symmetry if it is spontaneously broken
at a high scale by the VEVs of $T_3, \bar T_3, V_3, \bar V_3$. As we will
consider later, this can happen in two ways. One is by the hidden gauge group
(either $SU(5)_H$ or $SU(3)_H$) condensates $\langle V_3 \bar V_3 \rangle$,
$\langle T_3 \bar T_3 \rangle$ when these groups become strong. The other is by
explicit VEVs for any of the $V_3, \bar V_3, T_3, \bar T_3$ by the D
constraints for SUSY at M.
In both cases, the resulting axion, $a_3$, will be given by a linear
combination of the phases of {$V_3, \bar V_3, T_3, \bar T_3$} with nonzero
VEVs. If $V_3$ ($T_3$) gets
a VEV from $SU(3)_H$ ($SU(5)_H$) condensation, then $f_{a_3}=\langle V_3
\rangle
\sim \Lambda_3 \sim 10^{10}~GeV$ ($f_{a_3}=\langle T_3 \rangle \sim \Lambda_5
\sim 10^{14}~GeV$) for the hidden sector content of the model [\SLM,\EU]. If
$\langle V_3 \rangle$ or $\langle T_3 \rangle$ arise from F and D contraints,
then $f_{a_3} \sim {M/10} \sim 10^{17}~GeV$ which is much larger.

In order to solve the strong CP problem and be phenomenologically acceptable,
any of the three axions (or some linear combination of them) must satisfy
the following requirements.
a) The axion constant, $f_a$, has to be in the range $10^{10}~GeV<f_a<10^{12}~
GeV$ in order to satisfy the astrophysical and cosmological constraints.
b) The axion should remain massless up to QCD instanton effects. In particular,
it should not get mass from hidden sector gaugino condensation, trilevel and
non--renormalizable terms in the superpotential.
c) The axion should decouple from all observable neutral currents.

If these conditions are satisfied, then the effective $\theta$ is given by
$$\theta_{string}=a+\theta_{QCD}+\theta_{quark}+\theta_{gluino}+
\theta_{heavies} \eqno(8)$$
where $a$ is any one (or more) of the axions, the different $\theta$s are the
contributions of the QCD vacuum and the colored fermions: quarks, gluinos and
other heavy colored
fermions. In string models in the free fermionic formulation, gluino masses are
real and therefore $\theta_{gluino}=0$. In standard--like models there are
two colored heavy states $D_{45}$ and $H_{21}$ in the notation of Ref [11].
{}From the trilevel superpotential, we find that their masses can be made real
by choosing the VEVs of $\xi_1$, $\xi_3$ and $H_{18}$ real.
Then, we are left with the first three terms in Eq. (8) which is exactly the
usual $\bar \theta$ of the strong CP problem in QCD.

All of the axions above suffer from two problems: $f_a$s which are too large
and
couplings to the hidden sector gauge groups. $f_{a_1} \sim 10^{16}~
GeV$ as a result of the large compactification scale. $f_{a_2} \sim
M/10$ is fixed by the anomaly term in the D constraint given by Eq. (1). So,
$f_a \sim f_{a^{\prime}} \sim 10^{17}~GeV$ for the physical axions. For
$a_3$, $10^{10}~GeV<f_{a_3}<10^{14}~GeV$ depending on the detais of $U(1)_6$
breaking. It is well known that a large $f_a$ (i.e. $f_a>10^{12}~GeV$) means
that axion energy density $\rho_a>\rho_{crit}$ and therefore unacceptable
[\CH].

The three axions couple to the hidden sector gauge groups due to the anomalies
in those sectors (in addition to QCD). As a result, when these groups
become strong, gaugino condensates, ${\langle \tilde g \tilde g \rangle}$,
which
give masses to the axions form. The axion mass is [\KIM]
$$m_a^2 \sim {m_{\tilde g}{\langle \tilde g \tilde g \rangle} \over f_a^2}
\eqno(9)$$
where $m_{\tilde g}=\Lambda_s^3/M_P^2$ and $\langle \tilde g \tilde g \rangle=
\Lambda_s^3$. $\Lambda_s$ is the hidden gauge group condensation scale at which
supersymmetry is broken dynamically by gaugino condensates. Depending on
whether
$SU(5)_H$ or $SU(3)_H$ condenses, the physical axion masses are
$1~GeV<m_a \sim m_{a^{\prime}} <10^6~GeV$ and $10^5~GeV<m_{a_3}<10^{9}~GeV$.
Such heavy axions will decouple
from low energy QCD and will not solve the strong CP problem. (From the
cosmological point of view, they will decay very rapidly at early times and
will
not survive to play a part in QCD at later times.)

The $U(1)_A$ axion, $a_2$, also gets a mass from trilevel (and higher order
non--renormalizable) terms in the superpotential, $W_3$, since $W_3$ contains
terms of the form $\phi_1 \phi_2 \phi_3$ where all $\phi_i$ get VEVs and have
nonzero $Q_A$ charges. Then, $m_a \sim m_{a^{\prime}} \sim 10^{17}~GeV$ which
is much larger than gaugino induced masses. Therefore $a$ and $a^{\prime}$
decouple from the spectrum.

The $U(1)_6$ axion, $a_3$, on the other hand, does not get a mass from terms in
$W_3$ since none of the $V_3,\bar V_3, T_3, \bar T_3$ appear in $W_3$. In
addition, it does not get a mass from higher order non--renormalizable terms
either for the following reason. Any string of states which appear in $N>3$
terms of the superpotential must be neutral with respect to $Q_6$. If any one
or three of
{$V_3,\bar V_3,T_3,\bar T_3$} appear in the string to give mass to $a_3$,
then there must be another field with $Q_6 \not =0$ with nonzero VEV in the
string. But the other states with
$Q_6 \not =0$ are all in 16 of $SO(10)$ (or hidden sector states
with $Y \not =0$) and cannot get VEVs. The right--handed
neutrino, $N_3$, cannot get a VEV even though it is a Standard model singlet
in order not to generate a large $B-L$ violation at the scale $M$ [\SLM].
A string with two or four {$V_3,\bar V_3,T_3,\bar T_3$} is not possible due to
conservation of local $Q_1$ and $Q_2$. Therefore, $a_3$ gets a large mass only
from gaugino condensation in the hidden sector.

The axion $a_3$ has another desirable feature.
Since $V_3, \bar V_3, T_3, \bar T_3$ are neutral under $SU(2)_L$, $U(1)_L$
and $U(1)_C$, $a_3$ automatically decouples from the observable neutral
currents. In general, this decoupling requirement is
a constraint on the PQ charges of the states. In superstring models, since all
charges are already fixed, satisfying this constraint is not trivial.

\bigskip
\centerline {\bf 4. General results}

We saw that $a$ and $a^{\prime}$ both obtain large masses
from gaugino condensation in the hidden sector and have axion constants much
larger than cosmologically allowed. These results are completely general and
hold for all superstring models in the four dimensional free fermionic
formulation. Properties of $a_3$, on the other hand, depend on the details of
the specific model considered. In the model above, we saw that $a_3$ gets a
large mass from gaugino condensation whereas $f_{a_3}$ depends on the details
of spontaneous $Q_6$ breaking. We would like to know whether, in general, an
acceptable axion can arise in standard--like superstring models.
By acceptable we mean an axion without couplings to the hidden sector gauge
bosons and with $10^{10}~GeV<f_{a_3}<10^{12}~GeV$.

First, we show that the $Q_6$ color anomaly is a generic feature in models
with chiral families from the $Z_2$ twisted sectors and which use a $Z_4$
twist.
Inspecting the global $Q_6$ charges in $G_3$ and $Q_4,Q_5$ charges in $G_1,G_2$
we find that the $Q_6$ color anomaly arises from the ``wrong" sign of
$Q_6(d^c_3)={-1/2}$. This is a result of the GSO projection due to basis vector
$\gamma$. Before this projection, both signs for $Q_6(d^c_3)$ are possible but
$\gamma$ forces it to have the wrong sign. ($Q_6(d^c_3)=-Q_6(u^c_3)$ instead
of $Q_6(d^c_3)$=$Q_6(u^c_3)$). Note that $\gamma$ singles out a
generation by the number of periodic fermions in the three internal sectors.
The generation with the global $U(1)_{j+3}$ anomaly is the one corresponding
to ${1 \over 2}(b_j \cdot \gamma)=odd$. There is always one color
anomalous $U(1)_{j+3}$ which corresponds to the $b_j$ with ${1 \over 2}(b_j
\cdot \gamma)=odd$.
This follows from the fact that the right--handed part of $\gamma$ is
a $Z_4$ twist and there is always one internal sector with only
one periodic fermion in $\gamma$. For the model above, ${1 \over 2}(b_1,b_2)
\cdot \gamma=even$ and ${1 \over 2}(b_3 \cdot \gamma)=odd$ so the color anomaly
is in $U(1)_6$.

Even though the structure of $\alpha, \beta, \gamma$ is fixed, there is still
the freedom of permutations among them in the internal sector.
If by a permutation one gets ${1 \over 2}(b_j \cdot \gamma)=odd$ for $j \not
=3$
, then the
corresponding $U(1)_{j+3}$ will be color anomalous.  In addition, one can
reverse the signs of the coefficients $c\left(\matrix{b_j \cr \gamma \cr}
\right)$
from $-1$ to $1$. This does not affect the $U(1)_{j+3}$ color anomaly since
such a change reverses all global charges without changing their relative
signs.

A general property of models with a $Z_4$ twist is the presence of massless
sectors $b_j+ 2\gamma+(I)$ in the spectrum which contain
vector representations of the hidden non--Abelian gauge groups.
In order to remain light, $a_3$ should decouple from the hidden sector i.e.
it should be anomaly free
in the hidden sector. Is it possible to have a color anomalous global symmetry
without anomalies in the hidden sector? We will now show that the answer is
negative.

The hidden sector with the $U(1)_{j+3}$ anomaly is given by the sector $b_j+2
\gamma${\footnote*{In this section we write $b_j+2 \gamma$ for
$b_j+2 \gamma+(I)$ for notational simplicity.}}
with ${1 \over 2}((b_j+2 \gamma) \cdot \gamma)=odd$ as for $b_j$. When
${1 \over 2}((b_j+2 \gamma) \cdot \gamma)=odd$, $Q_{j+3}(T_j,V_j)=Q_{j+3}
(\bar T_j, \bar V_j)$ and there is a $U(1)_{j+3}$ anomaly in both $SU(5)_H$
and $SU(3)_H$ sectors. If
${1 \over 2}((b_j+2 \gamma) \cdot \gamma)=even$, then
$Q_{j+3}(T_j,V_j)=-Q_{j+3}(
\bar T_j, \bar V_j)$ and there is no anomaly in the hidden sector. Note that
the internal part of $b_j+2 \gamma$ is identical to that of $b_j$. As a result
of
this and the $Z_4$ twist in $\gamma$ , whenever ${1 \over 2}(b_j
\cdot \gamma)=odd$ for a particular $j$ (or generation), ${1 \over 2}((b_j+2
\gamma) \cdot \gamma)=odd$
for the same $j$. This means that whenever there is a color
anomalous global $U(1)_{j+3}$, the same $U(1)_{j+3}$ has anomalies also with
respect to the non--Abelian hidden gauge groups. We stress
that if $\gamma$ were a $Z_2$ twist like $\alpha$ and $\beta$, then ${1 \over
2}(b_j \cdot \gamma)=even$ and ${1 \over 2}((b_j+2 \gamma) \cdot \gamma)=odd$
for all $j$ and there would not be any anomalous global $U(1)$.

On the other hand, an acceptable axion constant, $10^{10}~GeV<f_{a_3}<
10^{12}~GeV$ can be obtained by $SU(3)_H$ (or $SU(3)$ subgroup of $SU(5)_H$)
condensation. Unfortunately this contradicts the requirements for realistic
quark mixing [\CKMM].
It can be shown that quark mixing arises due to VEVs of states from the sectors
$b_j+2 \gamma+(I)$. (This is another property of models with a $Z_4$ twist
$\gamma$.) In order to get realistic quark mixing, $V_j$ must get VEVs
$\sim 10^{16}~GeV$ which results in an unacceptably large axion constant.

\bigskip
\centerline {\bf 5. Conclusions}

There are three axions in standard--like superstring models. These are the
model
independent axion ($a_1$) which is present in all superstring models, the axion
of an anomalous $U(1)_A$ ($a_2$) which is present in all models in the free
fermionic formulation and the axion of a global $U(1)$ ($a_3$) which is model
dependent. $a_1$ and $a_2$ (or their linear combinations $a$ and $a^{\prime}$)
cannot solve the strong CP problem since they always
have large masses and axion constants. The model dependent axion, $a_3$, on
the other hand, can if it does not couple to the hidden sector and has an
acceptable constant. We show that in models with a $Z_4$ twist, there is always
a color anomalous global $U(1)$. Due to the $Z_4$ twist, the same global $U(1)$
necessarily has $SU(5)_H$ and $SU(3)_H$ anomalies in the hidden sector. In
addition, an acceptable axion constant is in conflict with requirements for
realistic quark mixing in these models. Therefore, $a_3$ cannot solve the
strong CP problem either.

In standard--like superstring models, another solution to the strong CP problem
which is vanishing $m_u$ is possible. This issue is currently under
investigation.

\bigskip
\centerline{\bf Acknowledgments}

I would like to thank Alon Faraggi for useful discussions. This work was
supported by a  Feinberg School Fellowship and the Department of Physics.

\refout

\vfill
\eject

\end

\input tables.tex
\nopagenumbers
%\special{landscape}
%\hoffset=1.25truein
%\nopagenumbers
\magnification=1000
%\font\normalroman=cmr10
%\font\style=cmr7
%\style
\tolerance=1200

%\fontdimen12\fivesy=0pt

%\textfont0=\sevenrm
%\scriptfont0=\fiverm
%\textfont1=\seveni
%\scriptfont1=\fivei
%\textfont2=\sevensy
%\scriptfont2=\fivesy

{\hfill
{\begintable
\  \ \|\ ${\psi^\mu}$ \ \|\ $\{{\chi^{12};\chi^{34};\chi^{56}}\}$\ \|\
{${\bar\psi}^1$, ${\bar\psi}^2$, ${\bar\psi}^3$,
${\bar\psi}^4$, ${\bar\psi}^5$, ${\bar\eta}^1$,
${\bar\eta}^2$, ${\bar\eta}^3$} \ \|\
{${\bar\phi}^1$, ${\bar\phi}^2$, ${\bar\phi}^3$, ${\bar\phi}^4$,
${\bar\phi}^5$, ${\bar\phi}^6$, ${\bar\phi}^7$, ${\bar\phi}^8$} \crthick
$\alpha$
\|\ 0 \|
$\{0,~0,~0\}$ \|
1, ~~1, ~~1, ~~0, ~~0, ~~0 ,~~0, ~~0 \|
1, ~~1, ~~1, ~~1, ~~0, ~~0, ~~0, ~~0 \nr
$\beta$
\|\ 0 \| $\{0,~0,~0\}$ \|
1, ~~1, ~~1, ~~0, ~~0, ~~0, ~~0, ~~0 \|
1, ~~1, ~~1, ~~1, ~~0, ~~0, ~~0, ~~0 \nr
$\gamma$
\|\ 0 \|
$\{0,~0,~0\}$ \|
{}~~$1\over2$, ~~$1\over2$, ~~$1\over2$, ~~$1\over2$,
{}~~$1\over2$, ~~$1\over2$, ~~$1\over2$, ~~$1\over2$ \| $1\over2$, ~~0, ~~1,
{}~~1,
{}~~$1\over2$,
{}~~$1\over2$, ~~$1\over2$, ~~0 \endtable}
\hfill}
\smallskip
{\hfill
{\begintable
\  \ \|\
${y^3y^6}$,  ${y^4{\bar y}^4}$, ${y^5{\bar y}^5}$,
${{\bar y}^3{\bar y}^6}$
\ \|\ ${y^1\omega^5}$,  ${y^2{\bar y}^2}$,
${\omega^6{\bar\omega}^6}$,
${{\bar y}^1{\bar\omega}^5}$
\ \|\ ${\omega^2{\omega}^4}$,  ${\omega^1{\bar\omega}^1}$,
${\omega^3{\bar\omega}^3}$,  ${{\bar\omega}^2{\bar\omega}^4}$  \crthick
$\alpha$ \|
1, ~~~0, ~~~~0, ~~~~0 \|
0, ~~~0, ~~~~1, ~~~~1 \|
0, ~~~0, ~~~~1, ~~~~1 \nr
$\beta$ \|
0, ~~~0, ~~~~1, ~~~~1 \|
1, ~~~0, ~~~~0, ~~~~0 \|
0, ~~~1, ~~~~0, ~~~~1 \nr
$\gamma$ \|
0, ~~~1, ~~~~0, ~~~~1 \|\
0, ~~~1, ~~~~0, ~~~~1 \|
1, ~~~0, ~~~~0, ~~~~0  \endtable}
\hfill}
\smallskip
\parindent=0pt
\hangindent=39pt\hangafter=1
%\normalroman
\baselineskip=18pt

{{\it Table 1.} A three generations ${SU(3)\times SU(2)\times U(1)^2}$
model. The choice of generalized GSO coefficients is:
${c\left(\matrix{b_j\cr
                                    \alpha,\beta,\gamma\cr}\right)=
-c\left(\matrix{\alpha\cr
                                    1\cr}\right)=
c\left(\matrix{\alpha\cr
                                    \beta\cr}\right)=
-c\left(\matrix{\beta\cr
                                    1\cr}\right)=
c\left(\matrix{\gamma\cr
                                    1,\alpha\cr}\right)=
-c\left(\matrix{\gamma\cr
                                    \beta\cr}\right)=
-1}$ (j=1,2,3),
with the others specified by modular invariance and space--time
supersymmetry. The twelve left--handed ($y^{1, \ldots, 6},\omega^{1, \ldots, 6}
$) and twelve right--handed ($\bar y^{1, \ldots, 6},\bar \omega^{1, \ldots, 6}$
) internal fermions which appear in complex or real pairs are divided into
three sectors, one for each generation.

\vfill
\eject

\end
\bye